\documentstyle[preprint,aps]{revtex}

\begin{document}
%%%  \draft \tighten
\title{ The $\beta$-delayed neutron emission in $^{78}$Ni region}
\author{I.~N.~Borzov}
\address {University of Leuven, B-3001 Leuven, Belgium}

\maketitle

\begin{abstract}
A systematic study of the total $\beta$-decay half-lives and $\beta$-delayed neutron emission probabilities is
performed. The $\beta$-strength function is treated within the self-consistent density-functional + continuum-QRPA
framework including the Gamow-Teller and first-forbidden transitions. The experimental total $\beta$-decay
half-lives for the Ni isotopes with $A\leq$76 are described satisfactorily. The half-lives predicted  from $A$=70
up to $A$=86 reveal fairly regular $A$-behaviour which results from simultaneous account for the Gamow-Teller and
first-forbidden transitions. For $Z\approx$ 28  nuclei, a suppression of the delayed neutron emission probability
is found  when the $N$=50 neutron closed shell is crossed. The effect originates from the high-energy
first-forbidden transitions to the states outside the $Q_{\beta} - S_n$-window in the daughter nuclei.
\end{abstract}

\section{Introduction}

Physics of the complex nuclear systems with high isospin asymmetry is a rich and multi-faceted field. It involves
such exotic objects like neutron stars and short-lived $\beta$-unstable nuclei with a large neutron (proton)
excess. The ground state and $\beta$-decay properties of these radioactive isotopes give a valuable information on
structural evolution far off stability and also provide input for supernova explosion calculations. In particular
for the modeling of the r-process nucleosynthesis, the most important are the nuclear masses which define the path
of the r-process through the neutron-rich domain of the nuclear chart. The $\beta$-decays and $\nu (\bar \nu)$
captures are also essential, as they  regulate the flow of the material to high $Z$-values and set up the r-process
time scale. The nuclei near the new doubly-magic $^{78}$Ni, $^{132}$Sn and "east" of $^{208}$Pb are of special
importance. Spectacular progress of the RNB experiments in these regions has been reached recently due to
$Z$-selective resonance-ionization laser ion-source  technique (see e.g \cite{Mish}). It has provided a unique
testing ground for theoretical approaches to exotic nuclei and modeling of explosive stellar events.

In this short paper we perform the microscopic study of the total $\beta$-decay half-lives and $\beta$-delayed
neutron emission probabilities of very neutron-rich nuclei. The  simultaneous analysis of these $\beta$-decay
observables gives a possibility to reconstruct the $\beta$-strength function. Specifically,  one may gain an
insight about the relative contribution of its Gamow-Teller (GT) and first-forbidden (FF) components. A
possibility of competition between the GT and FF decays in nuclei with the neutron-excess bigger than one major
shell has been often discussed in the literature (see \cite{KW} and Refs. therein). The experimental evidence of that
has been known since the first measurements in the region of $^{132}$Sn \cite{Fog}. However, the microscopic
global calculations \cite{KK,PM} have treated the total $\beta$-decay half-lives and delayed neutron emission
probabilities ($P_n$) in allowed transitions approximation. (The estimated contribution of the unique FF
transitions from \cite{KK98} has to be taken with some reservations.)

Clearly, the single $\beta$-decay channel approximation is not adequate for description of the isotopic dependence
of the $\beta$-decay characteristics, especially for the nuclei crossing the closed $N$ and $Z$ shells. It has been
pointed out recently in \cite{1B} that self-consistent framework of the finite Fermi system theory including the
GT and first-forbidden FF decays on the same footing allows for reasonably sound predictions of the ground state
properties and $\beta$-decay characteristics of very neutron-rich nuclei. An important contribution of the
first-forbidden (FF) non-unique decays to the total half-lives of neutron-rich nuclei above the closed proton and
neutron shells has been stressed in \cite{1B,3B}. The purpose of the present work is to examine the impact of the
high-energy FF decays on the $\beta$-delayed neutron emission in the nuclei near the closed shells at $Z$=28 and
$N$=50. In what follows we briefly outline the particulars of the theoretical framework, present new calculations
of the total $\beta$-decay half-lives and $\beta$-delayed neutron emission probabilities and discuss possible
uncertainties.

\section{Theoretical analysis}
The $\beta$-delayed neutron emission is basically a multi-step process consisting of
          a) the $\beta$-decay of the precursor ($A,Z$) which results in
             feeding  the excited states of the emitter nucleus ($A,Z+1$);
          b) neutron or/and $\gamma$-emission to an excited state
             or to the ground state of the final nucleus ($A-1,Z+1$).
The difference in the characteristic time-scales of the $\beta$-decay and subsequent particle emission
processes justifies an assumption of their statistical independence.
Thus, the total probability of the delayed neutron emission accompanying
the $\beta$-decay to the excited states in the daughter nucleus can be expressed
as \cite{PaSv}

\begin{equation}\label{1}
  P_n = T_{1/2} \int_{B_n(Z+1)}^{Q_{\beta}} {\rm d}{\omega} f_{0}(Z+1,\omega)
  \sum_{\beta} <\kappa_J> S_{\beta}(\omega,\gamma)P({j^{\pi}}_{i,f}, E_n),~
\end{equation}
\noindent where $\omega$ is the energy of nuclear transitions between the ground state of the
parent nucleus and final states of the emitter
\footnote{Within the self-consistent QRPA-approach one actually calculates the transition energy $\omega$
(clearly, an additional "outer" variable $Q_{\beta}$ \cite{KK} is needed to "derive" $E_x$).
The transition energies for the main GT and FF decays
reveal a smooth A-dependence which simplifies the analysis.}.
The  neutron emission threshold of the emitter $B_n(Z+1)$ and total $\beta$-decay energy
$Q_{\beta}$ are calculated self-consistently  \cite{BF}.

In the Eq.1,  $T_{1/2}$ stands for total $\beta$-decay half-life

\begin{equation} \label{2}
1/T_{1/2}= D^{-1} (G_A/G_V)^2
\int_0^{Q_{\beta}}{\rm d}{\omega} f_{0}(Z+1,\omega)
 \sum_{\beta} <\kappa_J> S_{\beta}(\omega,\gamma)~.
\end{equation}

Here the integrated lepton $(e^-,\bar \nu_e)$ phase-space
volume for allowed transitions  (the integrated relativistic Fermi function)
depending on the electron energy $W$
\begin{equation} \label{3}
f_0 = \int_{m_ec^2}^{W_{max}} F(Z+1,W)~p~W~(W_{max} - W)^2{\rm d}~W
\end{equation}
takes into account for the Coulomb and finite-size corrections \cite{GM}.

%%%%%%%%%%%%%%%%%%%%%%%%%%%%%%%%%%%%%%%%%%%%%%%%%%%%%%%%%%%%%%%%%%%%%%%%%%%%%%%
If the $\beta$-decay transition energy relative to the mother nucleus ground state satisfies $\omega <
Q_{\beta}-B_n \equiv Q_{\beta n}$, the neutron decaying state $|j^{\pi}_i \rangle$ locates within the so-called
$Q_{\beta n}$-window in the particle continuum; the energy of the emitted neutron being
$E_n=Q_{\beta}-B_n-\omega-E_{\gamma}$. The particle emission probability from the level $|j_i^{\pi_i}\rangle$ of
the emitter to the  level $|j_f^{\pi_f}\rangle$ of the final nucleus, as a function of the emitted neutron energy
$E_n$ is given by

\begin{equation}\label{4}
P({j{^\pi}}_{i,f}, E_n)=\frac{\Gamma_n(E_n)}{\Gamma_n(E_n)+\Gamma_{\gamma}(E_{\gamma})},~
\end{equation}
\noindent where $\Gamma_n({j^{\pi}}_{i,f}, E_n)$ is the neutron escape width and $\Gamma_{\gamma}({j^{\pi}}_{i,f},
E_{\gamma})$ - gamma ray width of the emitter. These can be expressed through the transmission coefficients for
l-wave neutron emission with the total angular momentum $j_n$ and corresponding $\gamma$-transmission coefficients
calculated from statistical model of Hausser-Feshbach. In what follows we will assume that the $\gamma$-emission
from neutron-unbound states is absent:  $\Gamma_{\gamma}\ll\Gamma_{n}$. The approximation simplifies the
calculations but  may cause some overestimation of the resulting $P_n$-values.

The intensity of feeding the excited states in the daughter nucleus is given by the set of $\beta$--strength
functions $S_{\beta}(\omega,\gamma)$ which describe the spectral distributions of the matrix elements of the GT
and FF $\beta$--decay transitions.  The index $\beta$ in the sum corresponds to the Gamow-Teller (GT) term with $L=0, J=1$ and non-unique
first-forbidden (FF) terms with $L=1, J=0,1$ treated in the $\xi$-approximation. The unique $J=2$ term can be also
retained but in the $\xi$-approximation it should be of minor importance. For the GT and non-unique FF decays
$<\kappa_{J=0,1}>$=1, and for the unique FF decays, $<\kappa_{J=2}>$=$f_1/f_0$, with the Fermi integral $f_1$
calculated as in \cite{Zyr}.

To calculate the $\beta$--strength functions  $S_{\beta}(\omega,\gamma)$  we use the continuum QRPA approach
based on the self-consistent ground state description within the local energy-density functional (DF)
theory. The Fayans phenomenological DF \cite{FTTZ00} consisting of a normal and a pairing part is adopted. The DF3
version of the functional \cite{BF} contains the two-body spin-orbit and velocity dependent effective
NN-interactions important for the full consistency, as well as the isovector spin-orbit force. The latter ensures
a correct description of the single-particle levels near the "magic-cross" at $^{132}$Sn \cite{Mez}. We  have
studied also the ground state description given by the Skyrme  MSk7 force \cite{MSk}. The
like-particle $T=1$ pairing in the ground state  is treated in the diagonal approximation on the basis confined by
the $E_{cutoff}$=15 MeV; the empirical strength of the $T=1$ pairing decreases slightly with increasing A.

The continuum QRPA (CQRPA) equations of the finite Fermi system theory are solved with an exact treatment of the
particle-hole (ph) continuum. The width $\gamma$ includes both the escape width  $\Gamma^{\uparrow}$ for the decay to the continuum, as
well as the spreading width of the isobaric states $\Gamma^{\downarrow}$. Within the CQRPA, the latter is assumed to depend linear on
the excitation energy \cite{Ber} (the cut-off at $\Delta \omega=2.5 \Gamma^{\downarrow}$ is employed for the
high-energy wings of the individual excitations with $\omega \leq Q_{\beta}$).

A finite-range effective NN-interactions in the particle-hole (ph) channel is chosen in  a $\delta+\pi+\rho$ form.
The one-$\pi$ and one-$\rho$ exchange terms modified by the nuclear medium are important in describing the
magnetic properties of nuclei and the nuclear spin-isospin responses. The competition between the one-pion
attraction $g_{\pi}Q<0$ and contact spin-isospin repulsion $g_{0}^\prime>0$, determines a degree of "softness" of
the pionic modes in nuclei that influences the $\beta$--decay half-lives. The set of the NN-interaction parameters
$g^{\prime}$=0.98, $g_{\pi}$=-1.38 has been used, corresponding to the quenching factor of the spin-isospin
response function $Q=e_q[\sigma\tau]^2$=0.81 (see \cite{1B}).

The effective $T=0$ NN-interaction in the particle-particle (pp) channel is chosen in the same $\delta$-function
form, as the like-particle $T=1$ pairing \cite {1B}. Neglecting the effective $T=0$ interaction would destroy the SO(8)
symmetry of the QRPA equations and cause unrealistic odd-even staggering of total $\beta$--decay half-lives.
Importantly, the CQRPA-like equations of the FFS allows for a reasonable description of nuclear spin-isospin modes
with practically A-independent constant of the pp-interaction which is far from the instability point in the pp
channel \cite{BFT90}.

\section{Results}

\subsection{The $\beta$-decay half-lives}
%and the Table
The total $\beta$-decay half-lives for Ni-isotopic chain  are displayed at Fig.1 in comparison to the finite-range
droplet model \cite{PM} and experimental data \cite{Fr98} - \cite{Am98}. The FRDM calculations overestimate the
experimental data and predict strong odd-even effect in the total $\beta$-decay half-lives. This  clearly
demonstrates a significance of the effective  pn-interaction in the particle-particle channel. Ignoring this
interaction violates the SO(8)-symmetry of the QRPA equations and leads to unrealistic scale of the odd-even
staggering of the total half-lives along the isotopic and isotonic chains.
The improved calculation performed in the present work is in better agreement with the experimental data for Ni
isotopes than the one from \cite{1B}. A slight underestimate of the experimental half-lives  \cite{Fr98}-\cite{Am98} is
observed for $A=74-76$. More importantly, the total half-lives in a rather long isotopic chain of $A=71-86$ reveal
a fairly regular behavior. Such a smooth A-dependence of the half-lives can be achieved only by taking into
account the forbidden decays. To stress this point, we show at the Fig.2 the isotopic dependence of the energies
for the main GT and FF transitions within the $Q_{\beta}$-window. Evidently, the FF  decays play  a  negligible role
for the nuclei with $A \leq 78$ because of the low transition energy (small available phase-space). In contrast,
after crossing the $N$=50 shell, the high-energy FF transitions give a dominant contribution to the total
half-life for the nuclei with $A \geq 79$.

For the Ga isotopes  with $A\le 84$ (Fig.3), both calculations give practically a similar agreement with the
experimental data \cite{Ree86} - \cite{Ru}. As seen from Fig.3, the FRDM+RPA calculations \cite{PM} underestimate
the experimental data for A=84, while our calculations overestimate them. Apparently, this is related to the
onset of the ground state deformation in the isotopes with $A\geq$ 84 which has not been included in our
calculations.

\subsection{The $\beta$-delayed neutron emission}
Within the allowed transitions approximation used in the global calculations \cite{KK,PM}, the isotopic dependence
of the $P_n$-value is rather schematic. It is mainly defined by the the decay to the so-called GT pigmy-resonance
located within the $Q_{\beta}$-window. The matrix element of the latter increase with $N$-$Z$, as well as the phase
space $Q_{\beta n}$ available for the delayed neutron emission. Thus, the $P_n$-value along the isotopic chain
simply tends to the maximum after the GT pigmy-resonance turns to be located within the $Q_{\beta n}$-window. On the
other hand, if the model includes the different $\beta$-decay channels, an isotopic dependence of the $P_n$-value is
defined by the relative energies of the GT and FF transitions which behave differently with increasing $A$.
Importantly, for the large enough neutron excess, the additional high energy FF transitions appear.

The calculations of the $\beta$-delayed neutron emission probabilities have been performed for several isotopic
chains with $Z\approx$28. Below, only the results  for Ni and Ga isotopic chain are discussed in detail.
As  can be seen from Fig.2,
for the Ni isotopes with $A\leq$79, no neutrons can be emitted following the decay to the pygmy GT resonance and
low transition energy FF states in the emitter nuclei. According to the present DF+CQRPA calculations they are
located higher than the $Q_{\beta n}$-window (Fig.2). For this reason, the increase of total delayed neutron
emission probability in the isotopes with A$\le$79 (Fig.4) is entirely due to relatively low-energy GT and FF
$\beta$-decays.

Starting from  $A$=79, the  GT pigmy-resonance enters the $Q_{\beta n}$-window (Fig.2). Assuming a pure GT-decay,
the $P_n$-values for  $A\geq$79 would tend to 100\%. However,  for the isotopes with $A\geq$79 crossing the $N$=50
shell, the drastic change in the relevant shell configurations takes place. From  Fig.2 which shows the  energies
of the different FF transition with the total transferred momentum $J=0$, it is seen that with filling of the $\nu
2 d_{5/2}$-orbital in the $A\geq$79 isotopes, the high-energy FF component of the $\beta$-strength function
appears due to the $\nu 2 d_{5/2} \rightarrow \pi 2 f_{5/2}$  transition. As mentioned above,  the high-energy FF
transitions give a dominant contribution to the total $\beta$-decay half-lives in the Ni isotopes with $A\geq$79.
At the same time, their calculated energies are higher than the $Q_{\beta n}$-threshold. Hence, the high-energy
part of the $\beta$-strength function  does not affect the delayed neutron emission, the feature that translates
into a "gap"-like pattern in the $P_n(A)$-curve (Fig.4). For the nuclei with $A\geq$86 the main  $J=0$ transitions
are located in the $Q_{\beta n}$-window (Fig.2), and the corresponding $P_n$-values tend to the maximum (Fig.4). It
would be important to check our prediction for Ni isotopic chain, though for $A\geq$80 for which the difference
with the GT approximation is observed it may be difficult at the moment.

In the context of the possible experimental measurements, the isotopes with higher $Z$ are of special interest. To
exemplify this we consider the case of Ga isotopes. For $^{80-81}$Ga, the FF transitions are of minor importance
but with increasing $A$ their contribution becomes very significant. It follows from Fig.5 that in contrast with Ni
isotopes, the $P_n$-values calculated for $^{82-87}$Ga isotopes are suppressed by the factor of 4-5 compared to the
ones corresponding to the GT approximation. Notice that in $^{82-84}$Ga, the evaluated  $P_n$-values \cite{Ru}
obtained by averaging of the existing experimental data \cite{Lu} - \cite{Ree85} can be described within the GT
approximation alone. However, it has to be realized that the discrepancy of the experimental data is very high.
Interestingly enough, in $^{83}$Ga, the ones from \cite{Ree85} are in agreement with the GT+FF calculations.
Assuming that reliability of the experimental data for  Ga isotopes has been questioned in \cite{Ru}, it would be
of great importance to perform the new measurements for this isotopic chain.

% L
% $\bullet$ Ge isotopes $P_n$ Fig.4

%%%%%%%%%%%%%%    %%%%%%%%%%%%     %%%%%%%%%%%%%
To conclude, one has to mention that in  Z=28 region, the delayed neutron emitter nuclei have relatively high
$Q_{\beta}$-values. For that reason, the accuracy  of the calculated $T_{1/2}$ and $P_n$-values is generally
higher than for the heavier nuclei with lower $Q_{\beta}$-values. However, as the delayed neutron emission is
typically a threshold phenomenon, it is of interest to analyze the different factors which may influence its
probability. The first concerns the nuclear pairing, as it changes the one-quasiparticle level energies and
calculated $B_n$-values. This issue has been tackled in \cite{Schuk}, however the only calculation was performed
for $^{137}$I with a very low $Q_{\beta n}$=1.6 MeV in which case a sensitivity to the paring strength is extremely
high. It would be of importance to study in detail how the predicted effect might be influenced by the different
prescriptions of nuclear pairing. To roughly estimate the possible sensitivity we varied the $B_n$-values within
the $\pm 1$ MeV. It turns out that for the nuclei with high $Q_{\beta n}$-values, the resulting impact factor is
not that strong e.g. for $^{83}$Ni the $P_n$-value varies within the  margins of 70.8 to 65.3$\%$ instead of
62.7$\%$ shown at Fig.4.

Second, the "gap-like" behavior of the $P_n$-values turns to be fairly robust against
the allowed choice of the effective NN-interaction parameters. (Using the set of the parameters corresponding to
the quenching factor of $e_q$=0.8 instead of 0.9 \cite{1B} leads to the  $P_n$-value of 71$\%$ for the $^{80}$Ni.)

Third, it is worth to estimate a possible influence of the $\beta$-strength redistribution due to the deformation
and/or the specific np-nh correlations  which may shift the $\beta$-strength outside or inside the $Q_{\beta
n}$-window. As it is seen from Fig.3 for the nuclei of interest beyond $^{80}$Ni, the GT pigmy-resonance energy is
much lower than the position of the $Q_{\beta n}$-window and such a shift seems quite unlikely. As for possible
impact of the deformation, the calculations \cite{MSk} predicts the quasi-spherical shape ($\beta_2 \leq$ 0.1) for
Ni isotopes with $A=74-94$ and for Ga isotopes with $A=79-83$.

Finally, as a proper account for the particle-emission channels within the Hausser-Feshbach framework has not been
performed, the predicted $P_n$-values may be considered as an upper limit estimate. Moreover, for very
neutron-rich nuclei the $B_n$-values become low, as well as the density of the neutron unbound levels in the near
threshold region. In such a case, a more detailed study of the $\beta$-delayed neutron emission may be of need
considering the contribution of direct neutron decay of the isobaric states.

\section{Summary}
The DF+CQRPA model of the delayed neutron emission is elaborated with the Gamow-Teller and first-forbidden
$\beta$-decay modes taken into account. The calculations of the total $\beta$-decay half-lives and delayed neutron
emission probabilities has been performed for nuclei near the closed shells at $Z$=28 and $N$=50. Within the
extended model, an agreement with the experimental data on the total half-lives is better than in our previous
calculations \cite{1B}. A suppression is predicted of the $\beta$-delayed neutron emission probabilities for $Z
\approx$28 isotopes above the $N$=50 shell. The first-forbidden decays in these nuclei are found to give a dominant
contribution to the total half-lives.  A threshold character of the delayed neutron emission makes it sensitive to
deblocking of the shell-configurations responsible for the high-energy forbidden $\beta$-decay transitions.
Inclusion of the FF $\beta$-decays leads to reduction of the total delayed neutron emission probabilities, as the
calculated  energies of the main FF transitions are systematically higher than the $Q_{\beta n}$-values.

For Ni isotopes, the scale of predicted reduction of the $P_n$-values  (compared to the ones estimated in the GT
approximation) is at the edge of the experimental precision achievable currently with the isotope-separated beams.
A similar "gap-like" behavior of the total probability of delayed neutron emission has been found in our
calculations for the  Cu isotopes. A stronger suppression of the $P_n$-values predicted for  Ga isotopes, as well as
for the Zn and Ge-As isotopic chains provides a certain challenge, as  these isotopes are more
accessible for the experimental measurements in the short term \cite{Fr04}.
%%%A  detailed calculations for a wide range of the nuclei near the neutron closed shells
%%%at N=50, 82, 126 has been done and will be reported elsewhere.

\section{Acknowledgements}
The author thanks the  OSTC, Belgium for  support within the PAI
Program IAP P5/07 "Exotic Nuclei for Nuclear Physics and Astrophysics".
Numerous discussions with M.~Arnould, S. Goriely, J.M.~Pearson, P.~Van Duppen,  M.~Huyse, D.~Pauwels,
N.~Severijns and J.-Ch.~Thomas are gratefully acknowledged.

\newpage
\begin{figure}
\caption{
The total $\beta$-decay half-lives for the
Ni-isotopes predicted from the FRDM+RPA  \protect\cite{PM}
and DF+CQRPA  \protect\cite{1B}; the experimental data are
taken from  \protect\cite{Fr98} -  \protect\cite{Am98}
}
%\label{T.Ni}
\end{figure}

\begin{figure}
\caption{
The calculated  position of the $Q_{\beta n}$-window for the delayed neutron
emission compared to the energies of the GT pigmy-resonance and main
$J^{\pi}=0^-$ transitions. For the lowest energy branch of the $0^-$ transitions
($\omega_1$), the points for A=77-81 are shown only
}
%\label{om.Ni}
\end{figure}

\begin{figure}
\caption{
The total $\beta$-decay half-lives for the
Ga-isotopes predicted from the FRDM+RPA  \protect\cite{PM}
and DF+CQRPA ( present work ); the experimental data are
taken from  \protect\cite{Ru}
}
%\label{T.Ga}
\end{figure}

\begin{figure}
\caption{
The total $\beta$-delayed neutron emission probabilities
for the Ni isotopic chain calculated within the DF+CQRPA
for the allowed GT transitions only ($\triangle$) and in the
GT+FF approximation ($\bigcirc$)
}
%\label{P.Ni}
\end{figure}

\begin{figure}
\caption{
The total $\beta$-delayed neutron emission probabilities
for the Ga isotopic chain calculated within the DF+CQRPA
for the allowed GT transitions only ($\triangle$) and in the
GT+FF approximation ($\bigcirc$). The experimental data are:
a) \protect\cite{Ru}, b) \protect\cite{Lu}, c) \protect\cite{Ree85}
}
%\label{P.Ga}
\end{figure}


\begin{references}

\bibitem{Mish}
V.I.~Mishin et al., Nucl. Instrum. Methods Phys. Res. {\bf B73}, 550 (1993);\\
P.~Van Duppen, ibid.,{\bf B73}, 66 (1997);\\
K.-L. Kratz, in ENAM 98, edited by B.M. Sherill et al., AIP Conf. Proc. No. {\bf 455}\\
(AIP , Woodbury, NY, 1998), p.827.
\bibitem{KW}
H.V. Klapdor, C.O. Wene, J. Phys. {\bf G6}, 1061 (1980).\
\bibitem{Fog}
J. Blomquist, M. Kerek, B. Fogelberg, Z.Phys. {\bf A314}, 199 (1983).\
\bibitem{KK}
M.~Hirsh, A.~Staudt, H.V.~Klapdor- Kleingrothaus, At.Data Nucl.Data Tables.  {\bf 51}, 244 (1992).\
\bibitem{PM}
P. Moeller et al.,  At. Data Nucl. Data Tables. {\bf 66}, 131 (1997).\
\bibitem{KK98}
H. Homma et al., Phys. Rev. {\bf C54}, 2972 (1996).\
\bibitem{BF}
I.N. Borzov et al., Z. Phys. {\bf A335}, 127  (1996).\
\bibitem{1B}
I.N. Borzov, Phys. Rev. {\bf C67}, 025802 (2003).\
\bibitem{3B}
H.~DeWitte et al, Phys. Rev. {\bf C69}, 044305 (2004).\
\bibitem{PaSv}
A. Pappas, T. Sverdrup, Nucl. Phys. {\bf A188}, 48 (1972).\
\bibitem{GM}
N.B. Gove, M.J. Martin,
 At. Data Nucl. Data Tables. {\bf 10}, 206 (1971).\
\bibitem{Zyr}
L.N. Zyryanova, {\it Unique beta transitions} (Russ. original, Acad.
Science USSR, Moscow-Leningrad, 1960); {\it Once forbidden Beta Transitions}
(translation Pergamon Press, 1963)
\bibitem{FTTZ00}
S.A. Fayans et al., Nucl. Phys. {\bf A676}, 49 (2000).\
\bibitem{Mez}
K.A. Mezilev et al., Pys. Scripta {\bf T56}, 272 (1975).\
\bibitem{MSk}
S.Goriely, F. Tondeur, J.M. Pearson, At. Data Nucl. Data Tables.  {\bf 77}, 311  (2000).\
\bibitem{Ber}
G.F. Bertsch, P.F. Bortignon, R. Broglia, Rev. Mod. Phys. {\bf 55}, 287 (1981). \
\bibitem{BFT90}
I.N. Borzov, E.L. Trykov, S.A. Fayans, Sov. J. Nucl. Phys. {\bf 52}, 627 (1990).\
%%%t1/2 Ni
\bibitem{Fr98}
S. Franchoo et al., Phys. Rev. Lett. {\bf 81}, 3100  (1998).\
\bibitem{Re85}
P.L. Reeder  et al., Phys. Rev. {\bf C31}, 1032  (1985).\
\bibitem{Am98}
F. Ameil et al., Eur. Phys.J. {\bf A1}, 1  (1998).\
% t1/2 Ga
\bibitem{Ree86}
P.L. Reeder et al., Research Rept. PNL-SA-14026 (1986).\
%%% t1/2  Pn    Ga
\bibitem{Lu}
E. Lund et al., Z. Phys. {\bf A294}, 233 (1980).\
\bibitem{Ru}
G. Rudstam, K. Aleklett, L. Sihver,
 At.Data Nucl.Data Tables. {\bf 53}, 1 (1993).\
\bibitem{Ree85}
P.L. Reeder et al.,
Proc. Am. Soc. Nucl. Chem. Meeting, Chicago (1985), p.171\
\bibitem{Fr04}
S. Franchoo, U. K\"oster, (private communication).\
\bibitem{Schuk}
D.S. Delion, D. Santos, P. Schuk, Phys. Lett. {\bf B398}, 1  (1997).\


\end{references}
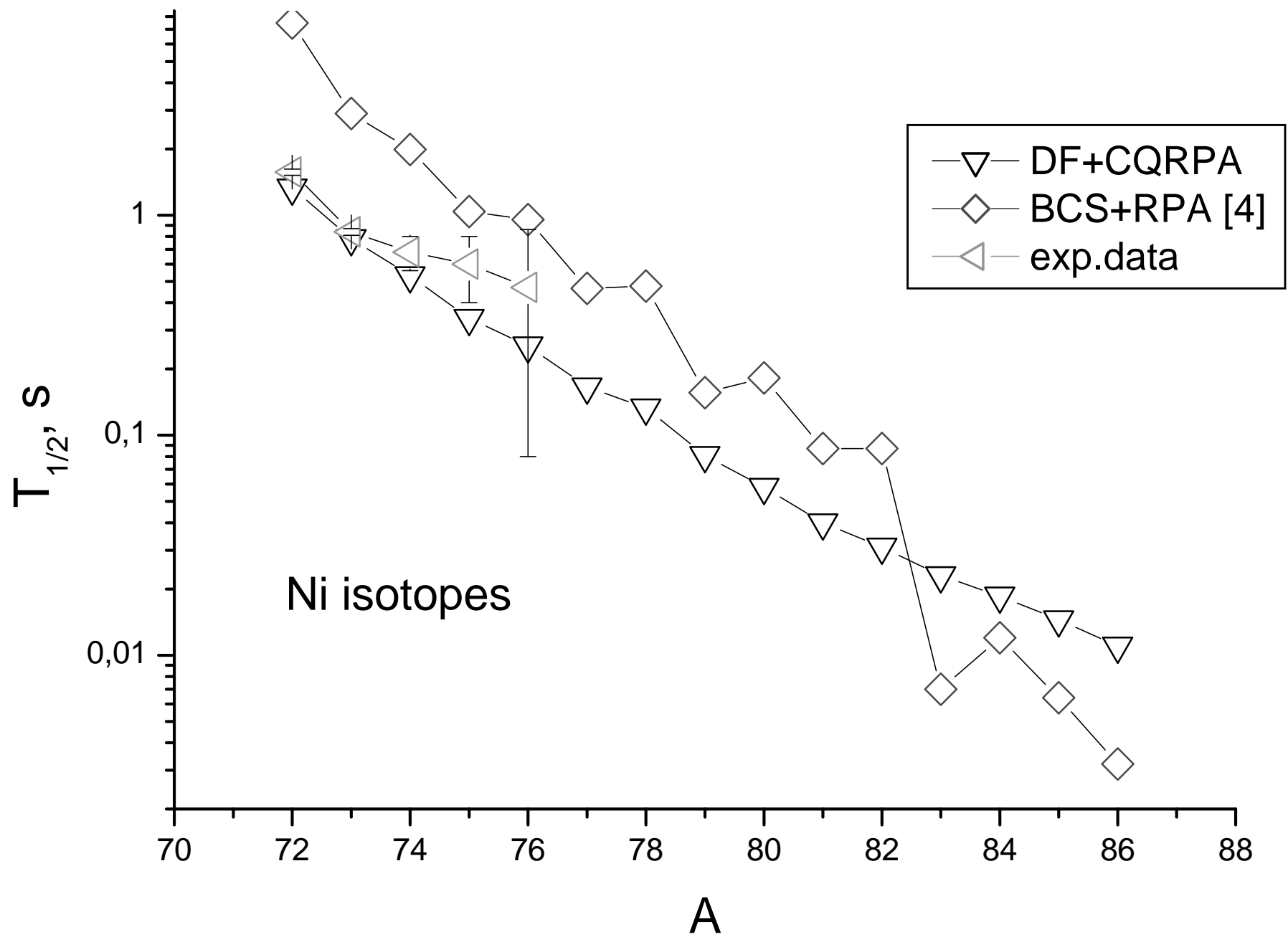
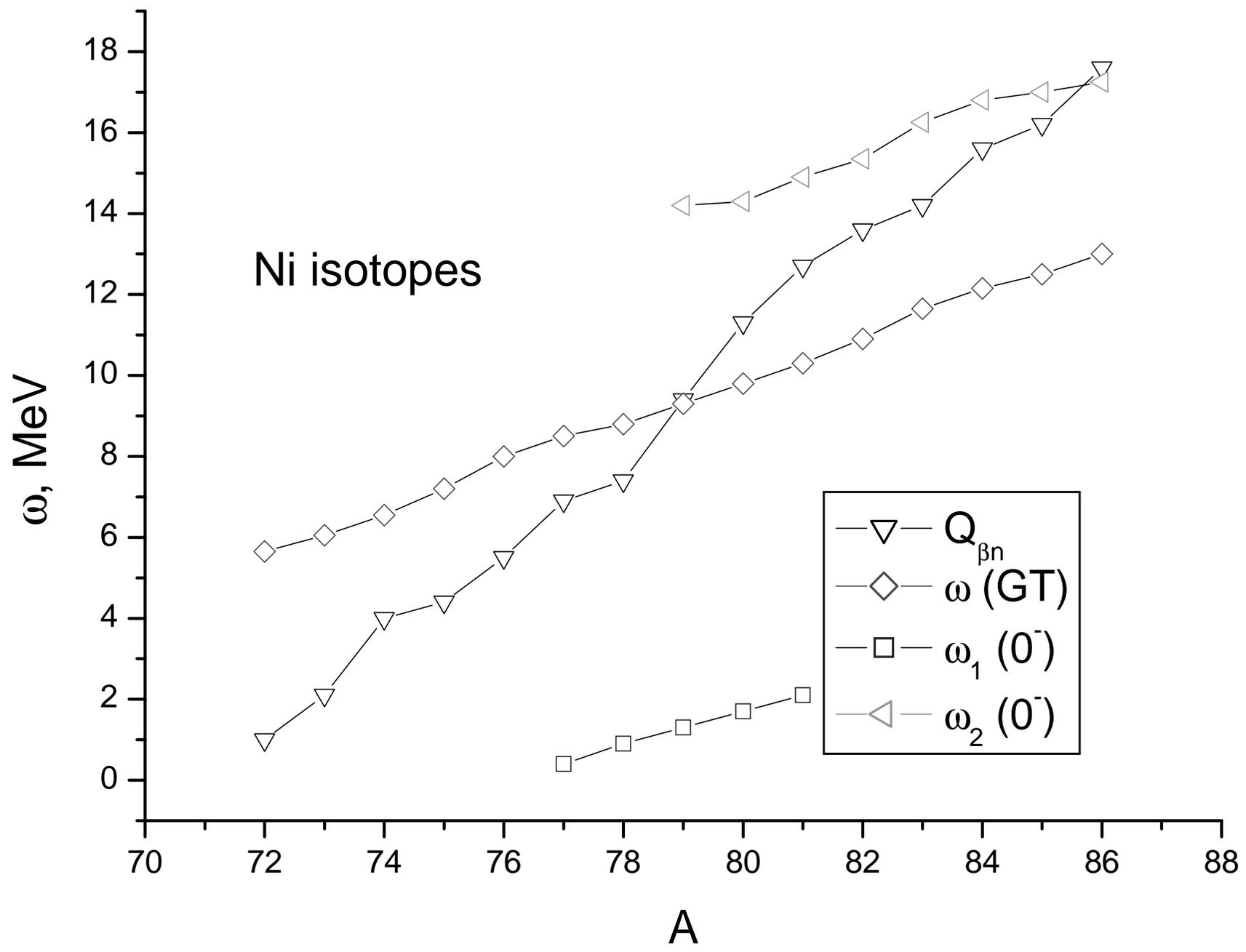
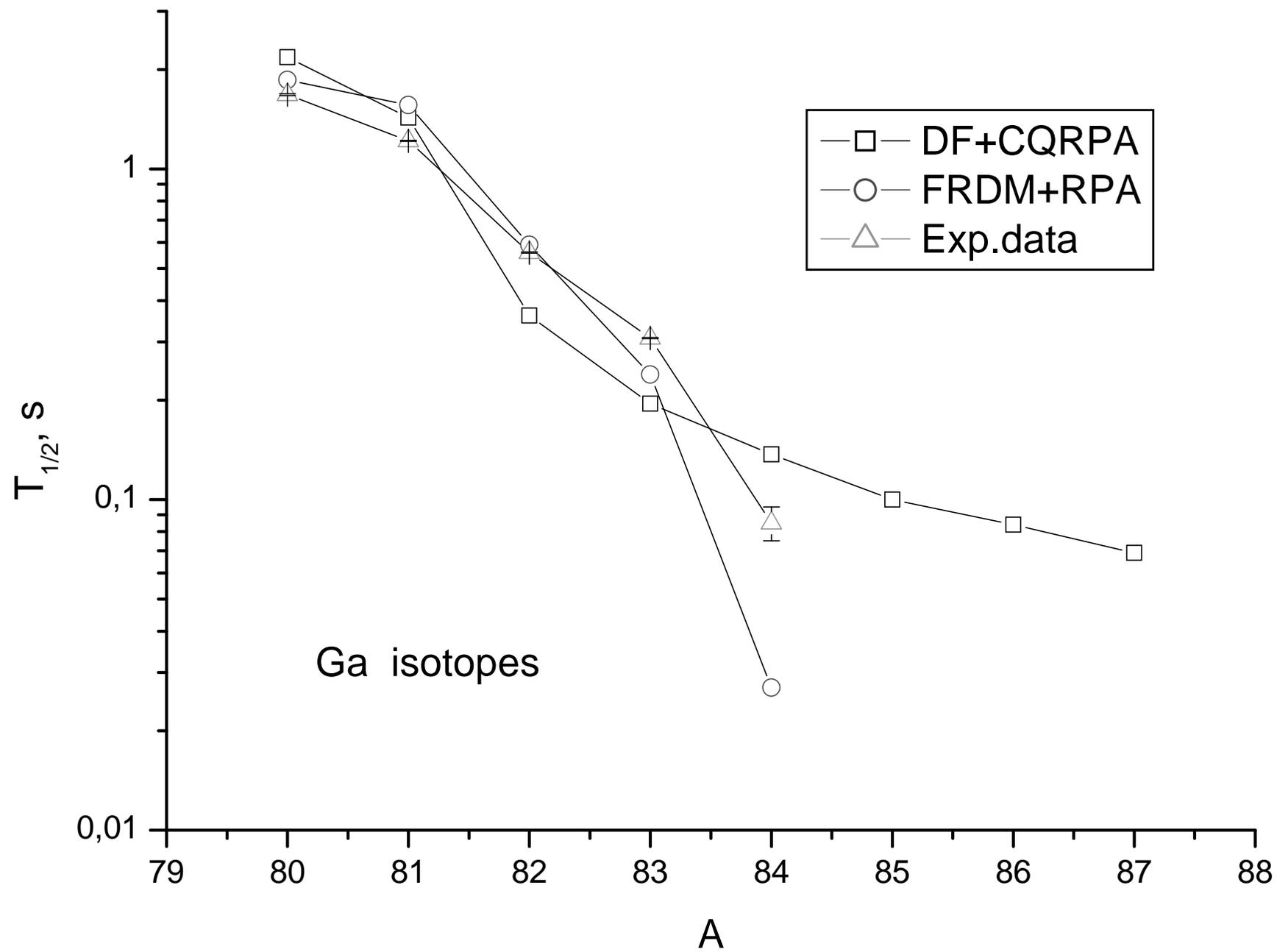
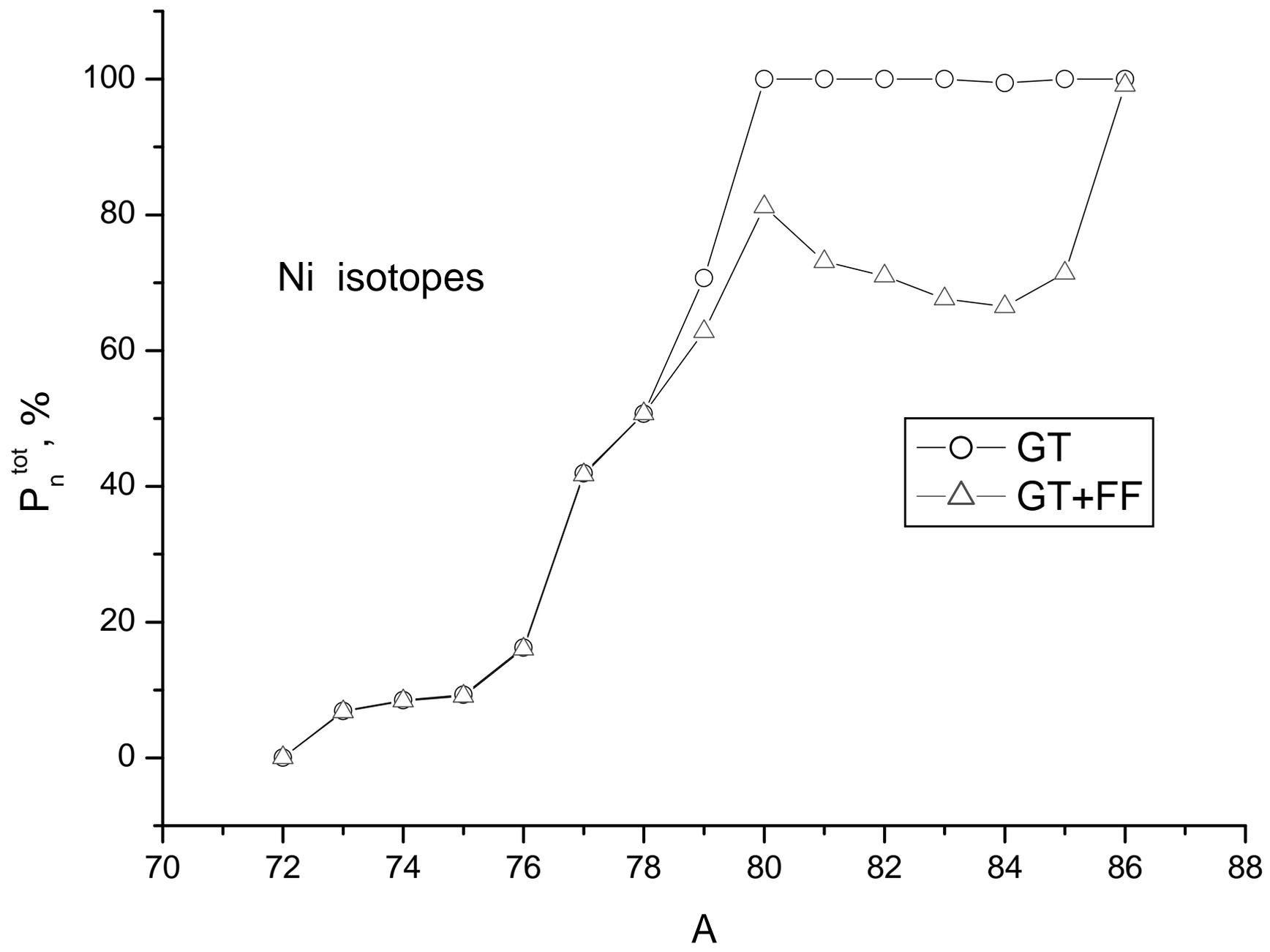
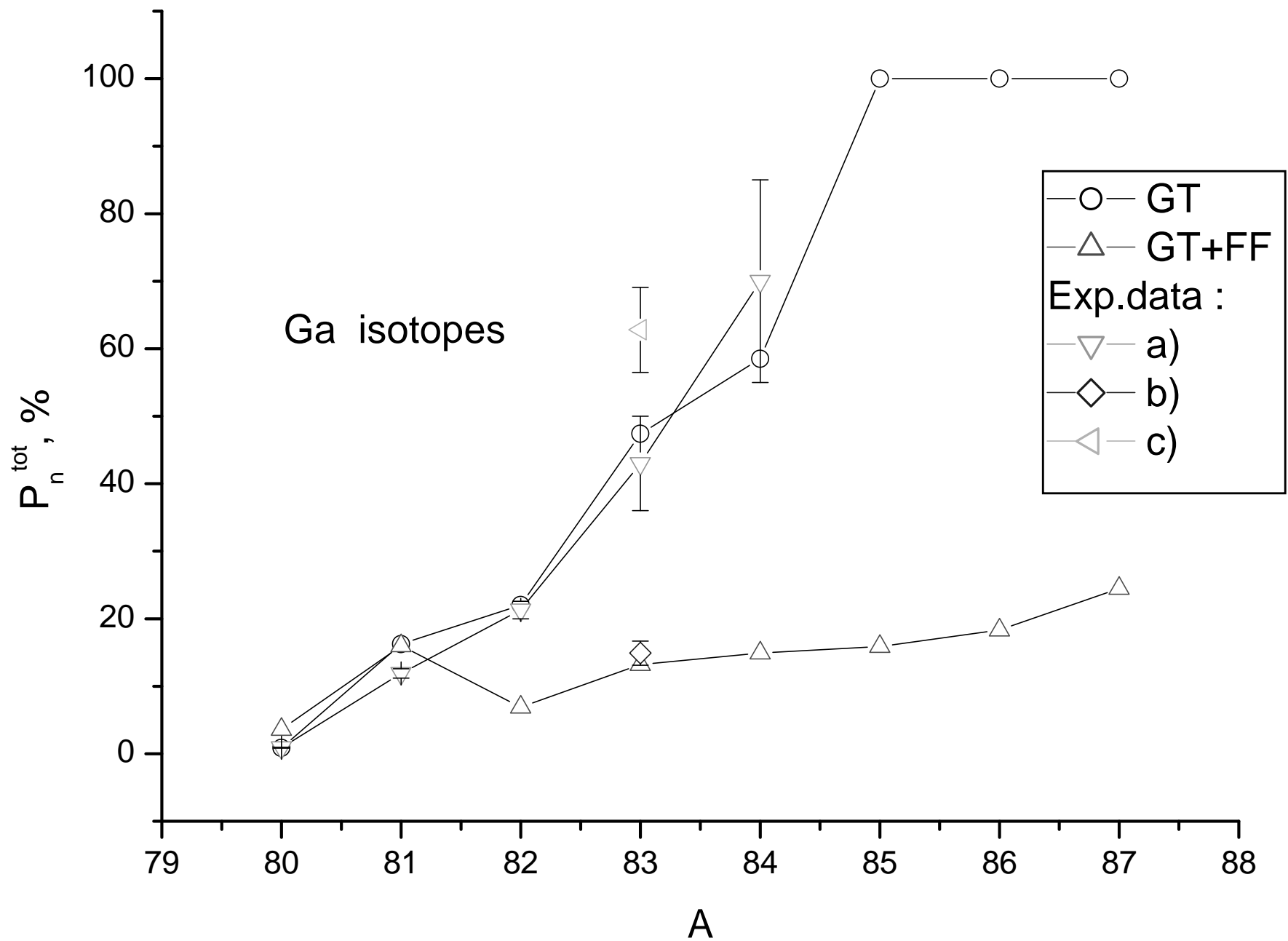
\end{document}